\documentclass[%
 reprint,
 amsmath,amssymb,
 aps,
]{revtex4-2}

\usepackage{graphicx}
\usepackage{dcolumn}
\usepackage{bm}
\usepackage{siunitx}

\begin{document}

\preprint{APS/123-QED}

\title{On the matrix formulation of time-delay interferometry}

\author{Jean-Baptiste Bayle}
\author{Michele Vallisneri}
\affiliation{Jet Propulsion Laboratory, California Institute of Technology, Pasadena CA 91109 USA}
\author{Stanislav Babak}
\affiliation{AstroParticule et Cosmologie (APC), Université de Paris/CNRS, 75013 Paris, France}
\affiliation{Moscow Institute of Physics and Technology, Dolgoprudny, Moscow region, Russia}
\author{Antoine Petiteau}
\affiliation{AstroParticule et Cosmologie (APC), Université de Paris/CNRS, 75013 Paris, France}

\date{\today}

\begin{abstract}
Time-delay interferometry (TDI) is a processing step essential for the scientific exploitation of LISA, as it reduces the otherwise overwhelming laser noise in the interferometric measurements. The fundamental idea, due to Armstrong, Estabrook, and Tinto [beginning with PRD \textbf{59}, 102003 (1999)], is to define new laser-noise--free observables by combining appropriately time-shifted measurements.
First- and second-generation TDI combinations cancel laser noise under the assumption that the LISA armlengths are constant or evolve linearly with time, respectively.
We recently extended TDI by solving for the laser-noise--free combinations \emph{implicitly}, writing the likelihood of the data directly in terms of the basic measurements, and using a discretized representation of the delays that can accommodate \emph{any} time dependence of the armlengths. We named the resulting formalism ``TDI-$\infty$'' [PRD \textbf{103}, 082001 (2021)].

According to Tinto, Dhurandhar, and Joshi [arXiv/2105.02054],
our matrix-based approach is invalidated by the simplified start-up conditions assumed for the design matrix that connects the time series of laser-noise fluctuations to the time series of interferometric measurements along the LISA arms.
Here we respond that, if those boundary conditions are indeed unrealistic, they do not invalidate the algorithm, since one can simply truncate the design matrix to exclude ``incomplete'' measurements, or set them to zero. Our formalism then proceeds unmodified, except that the length of the laser-noise canceling time series is reduced by the number of excluded measurements.
Tinto and colleagues further claim that the matrix formulation is merely a finite representation of the polynomial ring of delay operators introduced by Dhurandar, Nayak, and Vinet to formalize TDI [PRD \textbf{65}, 102002 (2002)]. We show that this is only true if all interferometric delays are exact multiples of the sampling interval, which will not be possible in practical contexts such as LISA.
\end{abstract}


\maketitle

\section{On boundary conditions}

In Ref.\ \cite{Vallisneri:2020otf}, we introduce TDI-$\infty$ using a toy model of the LISA measurements (see Fig.\ 1 therein). We consider a single laser noise source $\mathbf{c}$, split into beams that propagate along two arms, and are reflected back to the origin. Both beams interfere individually with the local laser source, yielding two sets of measurements. For brevity, we combine those measurements in a single vector $\mathbf{y}$. The design matrix $\mathsf{M}$ describes how laser noise appears in the measurements by way of $\mathbf{y} = \mathsf{M} \, \mathbf{c}$
%
%
(here we neglect any measurement noise). If we obtain $n$ measurement samples for each arm, then $\mathbf{c}$ and $\mathbf{y}$ have respective sizes $n$ and $2n$, and $\mathsf{M}$ is a $2n \times n$ matrix. TDI-$\infty$ constructs laser noise-free observables $\mathbf{t}$ as combinations of the measurements $\mathbf{y}$ such that $\mathbf{y}^\dagger \mathbf{t} = \mathbf{c}^\dagger \mathsf{M}^\dagger \mathbf{t} = 0$. In other words, TDI observables are identified as a vector basis for the null space of $\mathsf{M}^\dagger$.

Equation~(3) of Ref.\ \cite{Vallisneri:2020otf} exemplifies the design matrix for our toy model, assuming that one turns on the laser source at time $t = 0$, and that the propagation delays along both arms are constant multiples $2 \Delta t$ and $3 \Delta t$ of the sampling period.
Because of these delays, the first six measurements are \textit{incomplete}: the first lines of $\mathsf{M}$ contain exactly one non-vanishing coefficient.
Since $\mathsf{M}$ is full-rank, the size of the null space of $\mathsf{M}^\dagger$ (i.e., the number of TDI observables) is exactly $n$. (Note that TDI-$\infty$ does \emph{not} require that the two halves of $\mathsf{M}$, corresponding to measurements along the two arms, be separately invertible.)

Tinto \textit{et al.} note in \cite{Tinto:2021cwi} that such an $\mathsf{M}$ does not provide a realistic description of the measurements, thus invalidating our approach. That is because the incomplete measurements cannot be made in practice, given that no reflected beam is available to combine with the local laser.
While that is correct, physical boundary conditions can be obtained simply by truncating the design matrix to exclude the six incomplete measurements (or, equivalently, by setting them to 0). The vectors  $\mathbf{c}$ and $\mathbf{y}$ now have sizes $n$ and $2n-6$, and $\mathsf{M}$ is a $(2n-6) \times n$ matrix. The null space of $\mathsf{M}^\dagger$ continues to identify laser noise-free observables, with their number reduced by the number of excluded measurements. In our example, one finds $(2n - 6) - n = n - 6$ such observables. Thus, TDI-$\infty$ remains valid.


\section{On the matrix representation}

While the example discussed above assumes constant propagation delays that are multiple of the sampling cadence, the TDI-$\infty$ method can accommodate any fractional delays with any time dependences. This is emphasized in Ref.\ \cite{Vallisneri:2020otf} as one of the advantages of TDI-$\infty$, and an example is given in Fig.\ 2. To encode arbitrary delays in the design matrix, we employ finite-impulse-response fractional-delay filters based on $m$-point Lagrange interpolating polynomials. These filters act as approximations for the formal delay operators used in the classic formulation of TDI. Unphysical start-up conditions can be handled as described above, by excluding measurements until the fractional-delay filters ``clear'' the boundary.


Tinto and colleagues \cite{Tinto:2021cwi} claim that the matrix formulation is simply a representation of the polynomial ring of delay operators discussed by Dhurandhar, Nayak, and Vinet~\cite{Dhurandhar:2002zcl}, who identify TDI observables with elements in the module of polynomial syzygies.

However, the equivalence can be established only if propagation delays are constant multiples of the sampling interval, which would be impossible for LISA: measurements will be downlinked with $\sim 1$ Hz sample rates, whereas TDI requires delays accurate to $\sim 30$ ns.

The reason is that the set of fractional delay filters of length $m$ is not closed under composition. More precisely, the map $\varphi$ described in Eq.~(4.3) of~\cite{Tinto:2021cwi} is not a homomorphism: if $\ell_1$ and $\ell_2$ are two constant fractional propagation delays, the associated matrix representations $\varphi(\ell_1)$ and $\varphi(\ell_2)$ contain $m$ non-vanishing elements in each line, as does the matrix representation $\varphi(\ell_1 + \ell_2)$. However, the matrix composition $\varphi(\ell_1) \varphi(\ell_2)$ will contain as many as $3m - 2$ non-vanishing elements on each row, so $\varphi(\ell_1 + \ell_2) \neq \varphi(\ell_1) \varphi(\ell_2)$.  Practically, one would see for instance that building up Eq.\ (3.10) of Ref.\ \cite{Tinto:2021cwi} using matrix representations of the delays does \emph{not} solve $\mathbf{y}^\dagger \mathbf{t} = 0$ even for fixed non-integer-multiple delays. Thus, the matrix formulation of \cite{Vallisneri:2020otf} cannot be reduced to a representation of the polynomial ring of delay operators.

More importantly, unlike TDI-$\infty$ solutions, the elements of the module of polynomial syzygies cannot yield TDI observables that cancel laser noise for time-dependent armlength functions, simply because the polynomial-ring formalism does not ``know'' that the delays do not commute for time-dependent armlengths. That information is encoded in the design matrix $\mathsf{M}$, and TDI-$\infty$ uses it directly by identifying the combinations of interferometric samples that project out the laser noise.

The research was carried out at the Jet Propulsion Laboratory, California Institute of Technology, under a contract with the National Aeronautics and Space Administration (80NM0018D0004). Copyright 2021, California Institute of Technology.

\bibliographystyle{apsrev4-1}
\bibliography{references}

\providecommand{\noopsort}[1]{}\providecommand{\singleletter}[1]{#1}%
\begin{thebibliography}{3}%
\makeatletter
\providecommand \@ifxundefined [1]{%
 \@ifx{#1\undefined}
}%
\providecommand \@ifnum [1]{%
 \ifnum #1\expandafter \@firstoftwo
 \else \expandafter \@secondoftwo
 \fi
}%
\providecommand \@ifx [1]{%
 \ifx #1\expandafter \@firstoftwo
 \else \expandafter \@secondoftwo
 \fi
}%
\providecommand \natexlab [1]{#1}%
\providecommand \enquote  [1]{``#1''}%
\providecommand \bibnamefont  [1]{#1}%
\providecommand \bibfnamefont [1]{#1}%
\providecommand \citenamefont [1]{#1}%
\providecommand \href@noop [0]{\@secondoftwo}%
\providecommand \href [0]{\begingroup \@sanitize@url \@href}%
\providecommand \@href[1]{\@@startlink{#1}\@@href}%
\providecommand \@@href[1]{\endgroup#1\@@endlink}%
\providecommand \@sanitize@url [0]{\catcode `\\12\catcode `\$12\catcode
  `\&12\catcode `\#12\catcode `\^12\catcode `\_12\catcode `\%12\relax}%
\providecommand \@@startlink[1]{}%
\providecommand \@@endlink[0]{}%
\providecommand \url  [0]{\begingroup\@sanitize@url \@url }%
\providecommand \@url [1]{\endgroup\@href {#1}{\urlprefix }}%
\providecommand \urlprefix  [0]{URL }%
\providecommand \Eprint [0]{\href }%
\providecommand \doibase [0]{http://dx.doi.org/}%
\providecommand \selectlanguage [0]{\@gobble}%
\providecommand \bibinfo  [0]{\@secondoftwo}%
\providecommand \bibfield  [0]{\@secondoftwo}%
\providecommand \translation [1]{[#1]}%
\providecommand \BibitemOpen [0]{}%
\providecommand \bibitemStop [0]{}%
\providecommand \bibitemNoStop [0]{.\EOS\space}%
\providecommand \EOS [0]{\spacefactor3000\relax}%
\providecommand \BibitemShut  [1]{\csname bibitem#1\endcsname}%
\let\auto@bib@innerbib\@empty
\bibitem [{\citenamefont {Vallisneri}\ \emph {et~al.}(2021)\citenamefont
  {Vallisneri}, \citenamefont {Bayle}, \citenamefont {Babak},\ and\
  \citenamefont {Petiteau}}]{Vallisneri:2020otf}%
  \BibitemOpen
  \bibfield  {author} {\bibinfo {author} {\bibfnamefont {M.}~\bibnamefont
  {Vallisneri}}, \bibinfo {author} {\bibfnamefont {J.-B.}\ \bibnamefont
  {Bayle}}, \bibinfo {author} {\bibfnamefont {S.}~\bibnamefont {Babak}}, \ and\
  \bibinfo {author} {\bibfnamefont {A.}~\bibnamefont {Petiteau}},\ }\href
  {\doibase 10.1103/PhysRevD.103.082001} {\bibfield  {journal} {\bibinfo
  {journal} {Phys. Rev. D}\ }\textbf {\bibinfo {volume} {103}},\ \bibinfo
  {pages} {082001} (\bibinfo {year} {2021})},\ \Eprint
  {http://arxiv.org/abs/2008.12343} {arXiv:2008.12343 [gr-qc]} \BibitemShut
  {NoStop}%
\bibitem [{\citenamefont {Tinto}\ \emph {et~al.}(2021)\citenamefont {Tinto},
  \citenamefont {Dhurandhar},\ and\ \citenamefont {Joshi}}]{Tinto:2021cwi}%
  \BibitemOpen
  \bibfield  {author} {\bibinfo {author} {\bibfnamefont {M.}~\bibnamefont
  {Tinto}}, \bibinfo {author} {\bibfnamefont {S.}~\bibnamefont {Dhurandhar}}, \
  and\ \bibinfo {author} {\bibfnamefont {P.}~\bibnamefont {Joshi}},\
  }\href@noop {} {\  (\bibinfo {year} {2021})},\ \Eprint
  {http://arxiv.org/abs/2105.02054} {arXiv:2105.02054 [gr-qc]} \BibitemShut
  {NoStop}%
\bibitem [{\citenamefont {Dhurandhar}\ \emph {et~al.}(2002)\citenamefont
  {Dhurandhar}, \citenamefont {Rajesh~Nayak},\ and\ \citenamefont
  {Vinet}}]{Dhurandhar:2002zcl}%
  \BibitemOpen
  \bibfield  {author} {\bibinfo {author} {\bibfnamefont {S.~V.}\ \bibnamefont
  {Dhurandhar}}, \bibinfo {author} {\bibfnamefont {K.}~\bibnamefont
  {Rajesh~Nayak}}, \ and\ \bibinfo {author} {\bibfnamefont {J.~Y.}\
  \bibnamefont {Vinet}},\ }\href {\doibase 10.1103/PhysRevD.65.102002}
  {\bibfield  {journal} {\bibinfo  {journal} {Phys. Rev. D}\ }\textbf {\bibinfo
  {volume} {65}},\ \bibinfo {pages} {102002} (\bibinfo {year} {2002})},\
  \Eprint {http://arxiv.org/abs/gr-qc/0112059} {arXiv:gr-qc/0112059}
  \BibitemShut {NoStop}%
\end{thebibliography}%

\end{document}